\documentclass[conference]{IEEEtran}
\IEEEoverridecommandlockouts
\usepackage[superscript]{cite}
\usepackage{amsmath,amssymb,amsfonts}
\usepackage{graphicx}
\usepackage{booktabs}
\usepackage{textcomp}
\usepackage{url}
\usepackage{xcolor}
\newcommand{\hl}[1]{#1}



\usepackage{caption}
\def\BibTeX{\rmfamily B\kern-.05em\textsc{i\kern-.025em b}\kern-.08em T\kern-.1667em\lower.7ex\hbox{E}\kern-.125emX}

\begin{document}

\title{Can Large Language Models Reason about Event-Time Stream-Processing Semantics?}

\author{\IEEEauthorblockN{Zhuoxi Wang\textsuperscript{*}}
\IEEEauthorblockA{\textit{Northeastern University} \\
Boston, MA 02115, USA \\
wang.zhuox@northeastern.edu}
\and
\IEEEauthorblockN{Shibo Zheng}
\IEEEauthorblockA{\textit{Northeastern University} \\
Boston, MA 02115, USA \\
zheng.shib@northeastern.edu}
\and
\IEEEauthorblockN{Haoyu Zhang}
\IEEEauthorblockA{\textit{Northeastern University} \\
Boston, MA 02115, USA \\
zhang.haoyu6@northeastern.edu}}

\maketitle

\begin{abstract}
Streaming systems increasingly hand work to large language models (LLMs): writing pipelines, triaging alerts, reading logs. All of it assumes the model knows how event-time stream processing behaves, and we test that assumption directly. StreamReason-Bench asks a model to stand in for an event-time stream processor. Given a windowed query and a stream of out-of-order events, it reports which windows fire, with their aggregates, and which events are dropped as late. The answer key comes from a small reference implementation of Dataflow-model semantics, so we can grade exactly, and with a partial-credit row-F1, without running an engine. On 600 generated items covering tumbling, hopping, session, and processing-time windows, the models do poorly on event time. When told to answer directly, no model that actually follows the instruction clears 34\% exact match; chain-of-thought (CoT) roughly doubles that for several of them (GPT-4o goes from 0.34 to 0.48), and only one frontier model that reasons by default comes near solving the set (0.85). A processing-time control, with no watermarks and nothing late, is almost solved by every capable model. The gap points to event-time and late-data handling, not windowing or arithmetic, as the hard part. Sorting errors by window type tells the same story: late-data mistakes dominate the event-time windows and vanish on the control, while session windows mostly fail on where the session boundaries fall.
\end{abstract}


\section{Introduction}
Stream processing underpins real-time analytics, monitoring, and AIOps, and a growing body of work now puts LLMs inside these systems: turning natural language into streaming pipelines\cite{autostreampipe}, \hl{mediating anomaly detection over streams}\cite{calm}, and reading event logs\cite{loggpt,logprompt,logllm}. All of it takes for granted that the model knows how stream processing behaves. That behavior is not obvious, even to practitioners. What a query outputs depends on whether time is measured by event time or by arrival time, on the window type (tumbling, hopping, or session), \hl{on the watermark governing when each window is allowed to fire, and on the lateness rule determining which out-of-order events are still admitted}\cite{watermarks,streamsurvey}. Whether an LLM gets these rules right is, as far as we know, untested.

Benchmarks for LLMs on data tasks have so far concentrated on \emph{batch} text-to-SQL\cite{bird,spider2}, and they say almost nothing about streaming. The closest streaming work, AutoStreamPipe\cite{autostreampipe}, generates pipelines and scores them by whether they run without errors; it ships no benchmark of natural-language-to-query pairs and never checks whether the output is correct or whether the model grasps the semantics. Nothing yet measures whether an LLM can reproduce event-time behavior itself.

StreamReason-Bench is built around three choices. It is a benchmark, not a method, so our conclusions do not depend on the models doing well. It grades against a deterministic reference rather than noisy real-world labels, so a wrong answer is unambiguous. And it frames the task as reasoning, which is where LLMs are supposed to be strong.

We make three contributions. (1)~\textbf{StreamReason-Bench}, a benchmark of event-time stream-processing reasoning that needs no execution engine, with a self-tested reference executor for ground truth and a generator that scales across window types and difficulty levels. (2)~A two-protocol study (direct and CoT) of five LLMs: event time is hard, and the result depends heavily on how much the model reasons. No compliant model passes 34\% exact under direct prompting, and only the strongest reasoning model comes close under CoT. (3)~A processing-time control that pins the difficulty on event-time and late data, together with a failure-mode breakdown and ablations showing that CoT helps while a watermark scaffold and a worked example do not.

\section{Related Work}
\textbf{Text-to-SQL (batch).} \hl{BIRD}\cite{bird} \hl{asks} a model to \emph{generate} a relational query over static tables, and Spider~2.0\cite{spider2} pushes that to enterprise workflows but stays in the batch world. We instead ask the model to \emph{carry out} an event-time operator and report the result.
\textbf{LLMs for streaming systems.} AutoStreamPipe\cite{autostreampipe} writes Flink and Spark pipelines from natural language and scores them on whether they run, not on whether the answer is right; it releases no benchmark of query--output pairs. \hl{CALM}\cite{calm} \hl{puts an LLM judge inside a continuously adapting stream anomaly-detection loop, again assuming rather than testing the model's grasp of stream semantics.}
\textbf{LLMs for logs and AIOps.} LogGPT\cite{loggpt}\hl{,} LogPrompt\cite{logprompt}\hl{, and LogLLM}\cite{logllm} apply LLMs to log anomaly detection\hl{, and a recent systematic literature review maps this fast-growing line of work}\cite{logslr}. That line reads text lines and never touches windowing or watermark semantics.
\textbf{Stream-processing semantics.} \hl{Our reference implements the event-time semantics of the Dataflow model as realized in modern engines; the watermark machinery was pinned down precisely in a comparative study of Apache Flink and Google Cloud Dataflow}\cite{watermarks}\hl{, and a recent survey traces how production stream processors handle out-of-order data}\cite{streamsurvey}. All of it studies the semantics inside engines, not whether a language model can reproduce them.
\textbf{LLM reasoning.} Chain-of-thought prompting\cite{cot} and self-consistency\cite{selfconsistency} are the standard levers for multi-step reasoning, and we use CoT as one of our two protocols.
\textbf{Temporal reasoning.} LLMs are known to struggle with temporal reasoning\cite{tempreason,testoftime,timebench}. Our task is adjacent but distinct: it asks for the \emph{rule-governed} output of a streaming operator, not open-ended temporal inference.
\textbf{LLMs and time series.} A separate line reads raw numeric series with LLMs, for anomaly detection\cite{sigllm} and forecasting\cite{llmtime}, with mixed evidence on whether the LLM actually helps\cite{lmuseless}. We work on symbolic event-time operators rather than continuous signals.

\section{StreamReason-Bench}
\subsection{Task}
An item is a tuple $\langle \text{spec}, \text{stream}\rangle$. The \emph{spec} defines a windowing operator (type; size/hop/gap; aggregation $\in\{\textsc{sum},\textsc{count},\textsc{max}\}$; watermark delay; allowed lateness). The \emph{stream} is a list of events $\langle \text{key}, \text{value}, \text{event\_time}\rangle$ in \emph{arrival} order (event\_time may be out of order). The model outputs (a)~\textbf{emitted} windows---one row per (key, window) with the integer aggregate---and (b)~the \textbf{dropped} late events (arrival indices). We grade \emph{exact-match} (emitted set and dropped set both exactly correct) and \emph{row-F1} (set-F1 over emitted rows) for partial credit.

\hl{Row-F1 is computed per item as follows. Let $G$ be the set of gold emitted rows and $P$ the set of predicted rows, where a row is the complete tuple (key, window start, aggregate) for fixed windows and (key, session start, session end, aggregate) for sessions, and matching is exact on every field. Then}
\begin{equation}
\mathrm{prec} = \frac{|G \cap P|}{|P|},\quad
\mathrm{rec} = \frac{|G \cap P|}{|G|},\quad
\mathrm{F1} = \frac{2\,\mathrm{prec}\,\mathrm{rec}}{\mathrm{prec}+\mathrm{rec}}.
\end{equation}
\hl{Because the aggregate is part of the row identity, a window with the correct boundaries but a wrong aggregate contributes both a false positive and a false negative. The dropped-event set is graded only by exact match and never enters row-F1. An unparseable response scores 0, and an item whose gold output and prediction are both empty scores 1. Every reported row-F1 is the unweighted mean of per-item F1 (a macro average), and each per-window-type score reported in Section~\mbox{\ref{sec:results}} averages over the 150 items of that window type. Hopping windows deserve specific care because their windows overlap: with size $W$ and hop $H$, an event at time $t$ belongs to every window $[s, s{+}W)$ with $s$ a multiple of $H$ containing $t$, so the gold answer carries one row per covered (key, window start) pair. A model that emits rows for only one hop offset loses recall on the remaining offsets, and a model that misaligns window starts (for example, on multiples of $W$ rather than $H$) produces a near-disjoint row set and a near-zero item F1, so hopping row-F1 is deliberately sensitive to systematic offset errors while remaining blind to dropped-set errors, which only exact match captures.}

The three aggregations are chosen deliberately. \textsc{sum}, \textsc{count}, and \textsc{max} are all distributive, with a single integer of running state per window, so the arithmetic they demand is trivial and an error can be attributed to windowing, watermark, or late-data handling rather than to numeric computation. The near-perfect scores on the processing-time control, which uses the same aggregations \hl{(Section~\mbox{\ref{sec:results}})}, confirm that the arithmetic itself is not the obstacle. Holistic aggregates such as medians or percentiles, and richer stateful operators such as joins or pattern matching, would blur exactly this attribution by adding a second source of error; we leave them to the extensions discussed in Section~\ref{sec:discussion}.

\smallskip\noindent\emph{Worked example} (tumbling, $W{=}10$, \textsc{sum}, no watermark delay). Stream \texttt{[(a,1,0),(a,2,12),(a,9,3)]}: the watermark reaches 12 after the second event, so window $[0,10)$ has already closed by the time the third event arrives at \texttt{t=3}. The reference therefore emits \texttt{[[a,0,1],[a,10,2]]} and drops event \texttt{2}; the late value 9 never enters its window. Tracing this requires tracking the running watermark and each window's state at the same time, which is exactly where the models slip.

\subsection{Semantics (ground truth)}
We implement Dataflow-model semantics. After each arrival the watermark is
\begin{equation}
\mathrm{wm} = \max(\text{event\_time seen}) - \mathit{wm\_delay},
\end{equation}
and a window fires, emitting its aggregate and then closing, once
\begin{equation}
\mathrm{wm} \geq \text{window\_end} + \mathit{allowed\_lateness}.
\end{equation}
An event whose every containing window is already closed on arrival is dropped as late; at end-of-stream all open windows fire. Window types: \textbf{tumbling} $[\lfloor t/W\rfloor W, +W)$; \textbf{hopping} (size $W$, hop $H$): an event belongs to every window $[s,s{+}W)$ with $s$ a multiple of $H$ and $s\leq t<s{+}W$ (overlapping); \textbf{session} (gap $G$): per key, events within $G$ merge into one session, a bridging event merges two sessions, firing when $\text{watermark}\geq \text{end}+G+\textit{lateness}$; \textbf{processing-time (control)} (size $W$): assign by \emph{arrival position} $i\to[\lfloor i/W\rfloor W,+W)$, no watermarks, no late data. The reference is small and self-tested against hand-checked cases (7/7), including a subtle one where a watermark \emph{fires} an early session before a later bridging event can \emph{merge} it.

\subsection{Generation}
Items are generated programmatically (seeded) across difficulty tiers (easy/medium/hard) that scale stream length, number of keys, out-of-order degree, late-event rate, watermark delay, and allowed lateness; Table~\ref{tab:tiers} lists the tiers, and empty-output items are rejected. The tiers are designed so that each step up switches on one semantic mechanism rather than merely lengthening the stream. The easy tier is in-order and single-key with zero watermark delay, so it exercises only windowing and arithmetic; the medium tier adds out-of-order arrivals, a second key, and a nonzero watermark delay, so the model must track the watermark, although the late rate of 0 means no event can yet be dropped; the hard tier adds late events and a nonzero allowed lateness, engaging the full late-data machinery. Streams stay short (at most 8 events) at every tier, so failures reflect the semantics rather than context length. The monotone fall in accuracy from easy to hard for every model \hl{(Section~\mbox{\ref{sec:results}})} confirms that the ladder orders difficulty as intended. The released set has 600 items (4 window types $\times$ 3 difficulties $\times$ 50). Because gold is computed by the reference, the benchmark scales to thousands of items at no labeling cost.

\begin{table}[t]
\caption{Difficulty tiers and the parameters they scale, per item. $n$: stream length; keys: distinct keys; OoO: fraction of events perturbed out of order; late: late-event rate; $d_{wm}$: watermark delay; $L$: allowed lateness. Each (window type, tier) cell of the released set holds 50 items.}
\label{tab:tiers}
\centering
\begin{tabular}{lcccccc}
\toprule
Tier & $n$ & keys & OoO & late & $d_{wm}$ & $L$ \\
\midrule
easy   & 3--4 & 1 & 0.0 & 0.0 & 0    & 0    \\
medium & 4--6 & 2 & 0.4 & 0.0 & 0--5 & 0    \\
hard   & 5--8 & 2 & 0.7 & 0.3 & 0--5 & 0--5 \\
\bottomrule
\end{tabular}
\end{table}

\section{Experimental Setup}
Models: GPT-4o, GPT-4o-mini, Claude-Sonnet-4.6, Claude-Haiku-4.5, Gemini-2.5-Flash (API only, temperature~0). \hl{The evaluation framework itself needs no specialized hardware: all runs execute on a consumer laptop (Apple M4, 16\,GB RAM, macOS~26.5) under Python~3.9, and the reference implementation, item generator, and grader are pure Python with no GPU, no local model, and no training step. The models are reached exclusively through their public inference APIs via the official SDKs (\texttt{openai}~2.41.1, \texttt{anthropic}~0.109.1, \texttt{google-genai}~1.47.0), requesting the API model identifiers \texttt{gpt-4o}, \texttt{gpt-4o-mini}, \texttt{claude-sonnet-4-6}, \texttt{claude-haiku-4-5-20251001}, and \texttt{gemini-2.5-flash}; all calls were issued in June~2026 at temperature~0, so running time is bounded by API latency rather than by local compute, and anyone with API access can reproduce the leaderboard from the released harness.} Two \emph{protocols}: \textbf{direct} (``output only JSON'') and \textbf{CoT}\cite{cot} (``reason step by step, then output JSON''). Outputs are parsed by a balanced-brace extractor that takes the last JSON object containing \texttt{emitted} (robust to reasoning preambles). Completions are capped at 2000 output tokens. The cap budgets the reasoning trace, not the answer: no gold answer in the set exceeds 9 emitted rows or 5 dropped indices, which serializes to under 100 tokens, so at least 95\% of the budget is free for reasoning. We arrived at 2000 after an initial 700-token cap truncated the reasoning of the most verbose model. Truncation, when it occurs, surfaces as an unparseable response, because the extractor finds no complete JSON object; under the 2000-token cap, unparseable CoT responses number 13, 18, and 22 of 600 (2.2--3.7\%) for Claude-Sonnet-4.6, Claude-Haiku-4.5, and GPT-4o, so the cap is not a binding constraint on the reported scores. The two weakest CoT models leave more responses unparseable. For Gemini we cleared the cache and re-ran with the hardened parser and the scores did not change, so its collapse (Section~\ref{sec:results}) is not a truncation artifact; we cannot fully rule out budget exhaustion in GPT-4o-mini's unparseable output, but its exact-match sits near the floor under both protocols, so no conclusion rests on it. Predictions are cached for reproducibility.

\section{Results}\label{sec:results}
\begin{table}[t]
\caption{Main leaderboard ($n{=}600$): exact-match and row-F1 for each model under direct and CoT prompting. $^{*}$Sonnet-4.6 reasons by default, so its ``direct'' column is not a true direct condition. $^{\ddagger}$\hl{Gemini's CoT row-F1 falls because its hopping answers collapse, largely into responses with no extractable JSON; hardening the parser and re-running from a cleared cache left the scores unchanged, so this is the model's real CoT behavior, not an artifact of our extractor (see Failure Analysis).}}
\label{tab:leaderboard}
\centering
\begin{tabular}{lcccc}
\toprule
 & \multicolumn{2}{c}{Direct} & \multicolumn{2}{c}{CoT} \\
\cmidrule(lr){2-3}\cmidrule(lr){4-5}
Model & exact & row-F1 & exact & row-F1 \\
\midrule
Claude-Sonnet-4.6 & 0.85$^{*}$ & 0.97 & 0.77 & 0.94 \\
GPT-4o            & 0.34 & 0.72 & \textbf{0.48} & 0.79 \\
Claude-Haiku-4.5  & 0.29 & 0.62 & 0.47 & 0.77 \\
Gemini-2.5-Flash  & 0.32 & 0.63 & 0.41 & 0.46$^{\ddagger}$ \\
GPT-4o-mini       & 0.03 & 0.36 & 0.10 & 0.30 \\
\bottomrule
\end{tabular}
\end{table}

\begin{figure}[t]\centering
\includegraphics[width=\columnwidth]{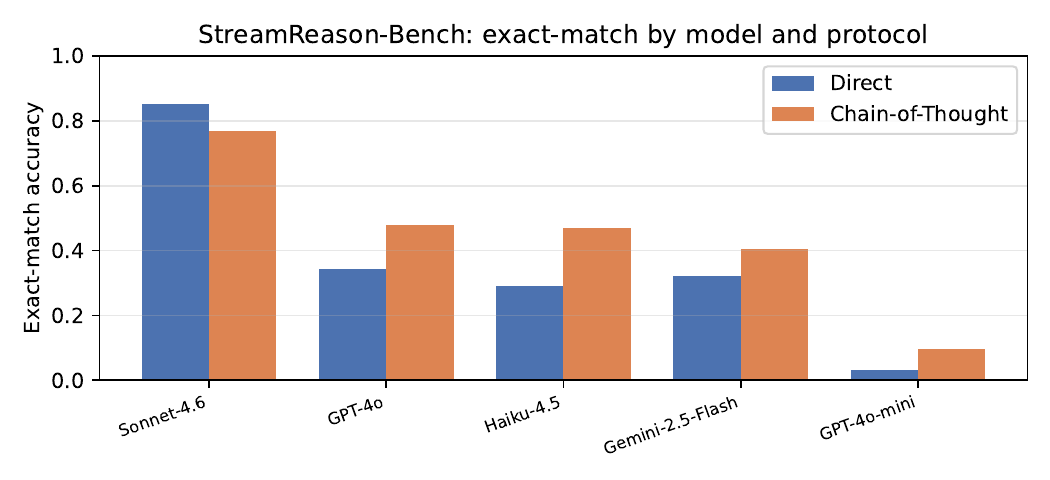}
\caption{Exact-match by model and protocol ($n{=}600$). Only the model that reasons by default clears 0.5; CoT lifts every model that follows the instruction, but none of the rest reach it.}\label{fig:leaderboard}
\end{figure}

Table~\ref{tab:leaderboard} and Figure~\ref{fig:leaderboard} summarize the runs. Two table footnotes are worth reading first. Sonnet reasons on its own even when told to answer directly, so its ``direct'' column is not really a direct condition ($^{*}$). And Gemini behaves oddly under CoT: exact match ticks up while row-F1 falls, because its hopping answers collapse\hl{, mostly into responses with no extractable JSON; we checked that this is the model's real CoT behavior, not an extractor artifact} ($^{\ddagger}$). The rest of the table is consistent. Answering directly is hard: no model that obeys the instruction exceeds 0.34 exact match. Chain-of-thought helps wherever the model follows it (GPT-4o rises from 0.34 to 0.48, Haiku from 0.29 to 0.47). And the set is far from solved. The only score above 0.5 belongs to the model that reasons by default (0.85), with every other model at or below 0.48 even with CoT.

\begin{table}[t]
\caption{Processing-time control vs.\ event-time windows (direct prompting, row-F1). The control uses the same windowing and arithmetic but has no watermarks and no late data, so the drop from the control column to the event-time columns isolates the difficulty.}
\label{tab:window}
\centering
\begin{tabular}{lcccc}
\toprule
Model & proc-time & tumbling & hopping & session \\
\midrule
Claude-Sonnet-4.6 & 1.00 & 0.97 & 0.94 & 0.97 \\
GPT-4o            & 0.90 & 0.70 & 0.66 & 0.60 \\
Gemini-2.5-Flash  & 0.90 & 0.76 & 0.62 & 0.24 \\
Claude-Haiku-4.5  & 0.82 & 0.70 & 0.58 & 0.37 \\
GPT-4o-mini       & 0.40 & 0.45 & 0.40 & 0.18 \\
\bottomrule
\end{tabular}
\end{table}

\begin{figure}[t]\centering
\includegraphics[width=\columnwidth]{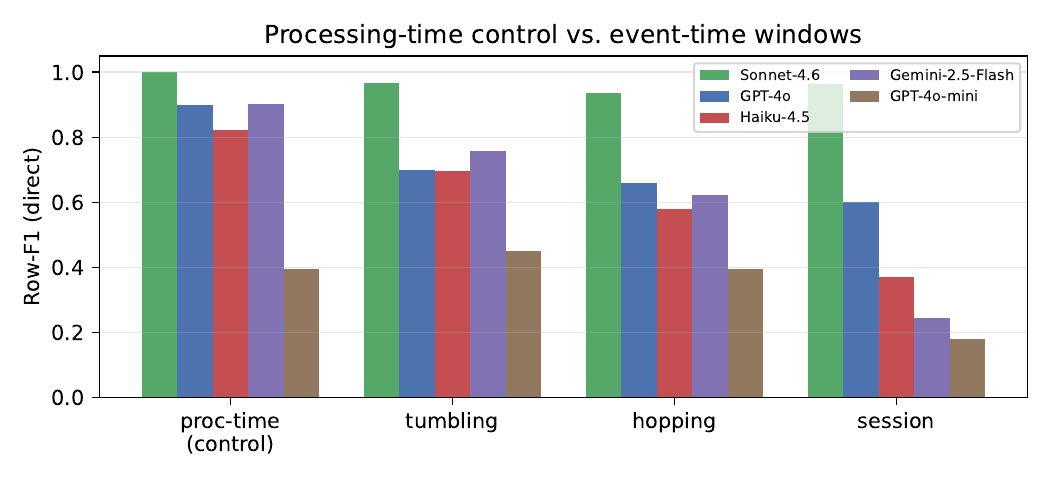}
\caption{Processing-time control vs.\ event-time windows (row-F1, direct). Every capable model is near-perfect on the control and clearly worse on the event-time windows, with session windows the hardest.}\label{fig:window}
\end{figure}

\begin{figure}[t]\centering
\includegraphics[width=0.85\columnwidth]{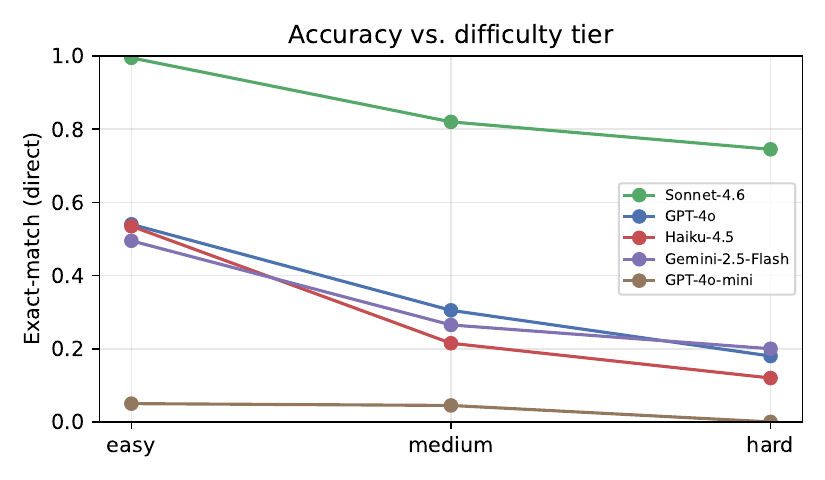}
\caption{Exact-match vs.\ difficulty tier (direct). Accuracy falls monotonically from easy to hard for every model, and the weaker ones bottom out near zero on the hard tier.}\label{fig:difficulty}
\end{figure}

\textbf{Processing-time control.} Table~\ref{tab:window} and Figure~\ref{fig:window} put the two regimes side by side. With processing time, where there are no watermarks and nothing arrives late, every capable model is close to perfect and well above its own event-time scores. Since the windowing and the arithmetic are the same in both regimes, the gap has to come from event time and late data. Scores also fall steadily from the easy tier to the hard one (Figure~\ref{fig:difficulty}, with exact values in Table~\ref{tab:difficulty}): even Sonnet drops from 0.99 to 0.74, and the weaker models flatten near zero on the hard tier.

\begin{table}[t]
\caption{Exact-match by difficulty tier (direct prompting). The tiers scale stream length, keys, out-of-order rate, late rate, watermark delay, and allowed lateness, as defined in Table~\ref{tab:tiers}.}
\label{tab:difficulty}
\centering
\begin{tabular}{lccc}
\toprule
Model & easy & medium & hard \\
\midrule
Claude-Sonnet-4.6 & 0.99 & 0.82 & 0.74 \\
GPT-4o            & 0.54 & 0.30 & 0.18 \\
Claude-Haiku-4.5  & 0.54 & 0.21 & 0.12 \\
Gemini-2.5-Flash  & 0.49 & 0.27 & 0.20 \\
GPT-4o-mini       & 0.05 & 0.04 & 0.00 \\
\bottomrule
\end{tabular}
\end{table}

\section{Failure Analysis}
\begin{figure}[t]\centering
\includegraphics[width=\columnwidth]{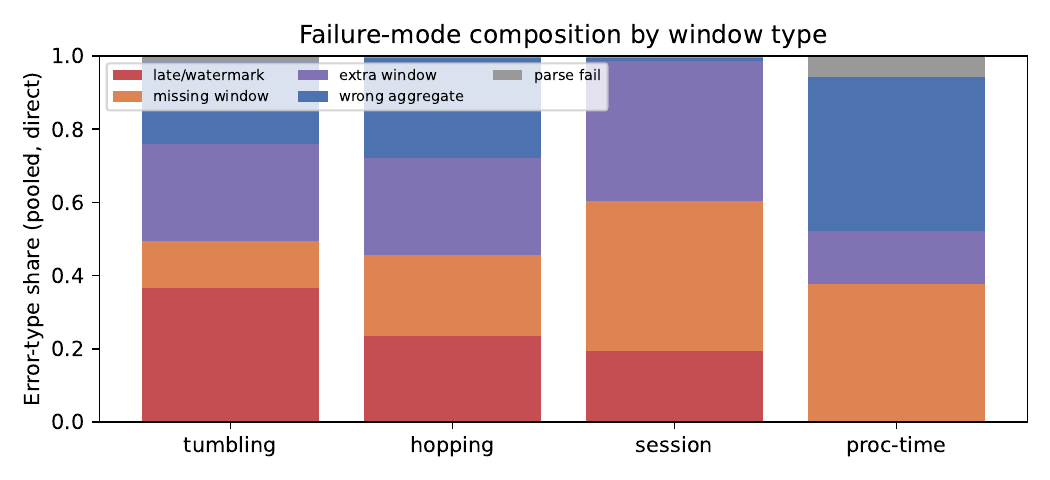}
\caption{Failure-mode composition by window type (pooled across models, direct prompting). Late-data errors dominate the event-time windows but vanish for the processing-time control, which has no watermarks; session windows fail through merge boundaries, i.e.\ missing and extra windows; hopping tilts toward aggregation errors.}\label{fig:failures}
\end{figure}

We label every imperfect answer by error type. \hl{The labeling is programmatic and deserves a precise statement, in particular for unparseable output. An answer is categorized \emph{unparseable} if and only if the balanced-brace extractor finds no complete JSON object containing an \texttt{emitted} field anywhere in the response; such answers receive that single exclusive label, and we make no attempt to reconstruct intent from surrounding prose (they also score 0 exact and 0 row-F1, so the failure figure and the leaderboard treat them consistently). Every other imperfect answer is parsed, and receives one or more of four tags computed against gold: \emph{missing window} (a gold window identity absent from the prediction) and \emph{extra window} (a predicted identity absent from gold), where the identity is the row without its aggregate, i.e., (key, window start) for fixed windows and (key, start, end) for sessions; \emph{wrong aggregate} (a shared identity whose values disagree); and \emph{late-data error} (predicted dropped set differs from gold). Tags are not mutually exclusive, so Figure~\mbox{\ref{fig:failures}} reports the share of tag occurrences within each window type, pooled over the five models under direct prompting. One residual pattern, an answer that emits several rows for the same window identity (typically a wrong and a right aggregate for the same window) and therefore fails exact match while triggering no tag, accounts for at most 2\% of items for any model and is tracked in the released harness but excluded from the figure.} The mix of error types looks about the same from one model to the next, so a per-model split says little; a split by window type (Figure~\ref{fig:failures}, pooled over models) is more telling. Late-data and watermark mistakes are the largest category on tumbling windows (about 32\%), and on the event-time windows generally, but they disappear on the processing-time control, which has no watermarks to get wrong. Session windows fail somewhere else entirely: missing and extra windows together make up about 71\% of their errors, while wrong aggregates barely register (around 1\%), so the trouble is deciding where one session ends and the next begins. Hopping windows tilt back toward aggregation errors, which fits their overlapping assignment. \hl{Unparseable output is rare under direct prompting: 0, 1, 1, 10, and 42 responses of 600 for Claude-Sonnet-4.6, GPT-4o, Gemini-2.5-Flash, Claude-Haiku-4.5, and GPT-4o-mini, so it never dominates the direct-protocol failure mix in Figure~\mbox{\ref{fig:failures}}. Under CoT it grows (13, 22, 269, 18, and 255 respectively, same model order): for Gemini-2.5-Flash and GPT-4o-mini the CoT collapse is dominated by responses that contain no complete JSON object at all, and these enter the breakdown as unparseable, not as guessed error types. We verified that this is the models' real CoT output behavior rather than an artifact of our extractor: hardening the parser and re-running Gemini from a cleared cache left its scores unchanged (Section~\mbox{\ref{sec:results}}).}

\section{Ablations}
\begin{table}[t]
\caption{Prompt ablations on event-time items (exact-match). Of three interventions on the two GPT models, only chain-of-thought moves the numbers; a watermark scaffold and a one-shot demonstration do not.}
\label{tab:ablation}
\centering
\begin{tabular}{lcc}
\toprule
Intervention & GPT-4o-mini & GPT-4o \\
\midrule
base (direct) & 0.00 & 0.20 \\
chain-of-thought & \textbf{0.13} & \textbf{0.39} \\
watermark trace (scaffold) & 0.01 & 0.21 \\
one-shot demonstration & 0.02 & 0.19 \\
\bottomrule
\end{tabular}
\end{table}
Of the three interventions, only chain-of-thought helps (Table~\ref{tab:ablation}). Handing the model the watermark value after each event does not, and is in fact a touch worse; a single worked example does not help either. If the problem were not knowing how to compute a watermark, or how to format the output, one of these would have moved the numbers. What is left is the step-by-step bookkeeping of the simulation itself, which is exactly what CoT gives the model room to do.

\section{Discussion and Limitations}\label{sec:discussion}
\textbf{Is the answer key right?} It comes from our reference implementation, so we check it in two ways: against hand-worked cases (7 of 7) and against a separate re-implementation run on all 300 fixed-window items, which agree completely. The second check paid off. It caught a non-standard late-data case in an earlier version, where a window would reopen for a late first element, and we fixed it before reporting any numbers. We still plan to cross-check a sample against a production engine such as Beam or Flink. \textbf{How much does the protocol matter?} Quite a bit: both the amount of reasoning we allow and the output-token budget move the scores, so we fix them and report direct and CoT next to each other. CoT is not a guaranteed win either, as Gemini's row-F1 shows. \textbf{Scope.} We stay with univariate streams, \textsc{sum}/\textsc{count}/\textsc{max}, and four window types; interval joins and more elaborate watermark strategies are left for later, as are holistic aggregations (medians, percentiles) and multi-stream operators, which would test state management as much as event-time semantics.

\textbf{What this is for.} The practical reading is a cautious one. An LLM asked to reason about event-time behavior, to explain why a windowed query returned what it did or to predict its output, should not be taken at its word without either chain-of-thought or an external check, and a larger model is not automatically the safer choice: Gemini regresses under CoT, and the mini model is wrong almost everywhere. For research, the benchmark is a cheap and unsaturated target. The gap between the processing-time control and the event-time windows is a clean handle on temporal bookkeeping specifically, and the generator can dial up difficulty on demand. The natural next steps are interval and multi-stream joins, longer streams that tax the model's working memory, and methods that let the model \emph{offload} the bookkeeping, by writing and running code or calling a stream engine as a tool, instead of carrying it out in its head. That last route is, in the end, what production systems already do; whether an LLM can drive that machinery correctly is the question our results leave open.

\textbf{Reproducibility.} The 600 items, the reference implementation, the generator, and the evaluation harness, with both protocols and cached model outputs, are released so the numbers can be reproduced and the benchmark extended.

\section{Conclusion}
Today's LLMs handle event-time stream semantics badly, watermark-gated firing and late data in particular. Most models fail when asked to answer directly, chain-of-thought is needed to make any real progress, and only the strongest reasoning model gets close, even though the processing-time version of the same task is easy for almost everyone. StreamReason-Bench is cheap to run, reproducible, and nowhere near saturated, which makes it a concrete target for the temporal reasoning these models will need as they move deeper into streaming systems.

\end{document}